\documentclass[reprint,superscriptaddress,amsmath,amssymb,aps]{revtex4-1}
\usepackage{graphicx}
\usepackage{bm}

\begin{document}

\title{Externally Controlled Lotka-Volterra Dynamics \\ 
in a Linearly Polarized Polariton Fluid}

\author{Matthias Pukrop}
\affiliation{%
 Department of Physics and CeOPP, Universit\"at Paderborn, Warburger Stra{\ss}e 100, 33098 Paderborn, Germany
}
\author{Stefan Schumacher}
\affiliation{%
 Department of Physics and CeOPP, Universit\"at Paderborn, Warburger Stra{\ss}e 100, 33098 Paderborn, Germany
}%
\affiliation{College of Optical Sciences, University of Arizona, Tucson, AZ 85721, USA}

\date{\today}

\begin{abstract}
Spontaneous formation of transverse patterns is ubiquitous in nonlinear dynamical systems of all kinds. An aspect of particular interest is the active control of such patterns. In nonlinear optical systems this can be used for all-optical switching with transistor-like performance, for example realized with polaritons in a planar quantum-well semiconductor microcavity. Here we focus on a specific configuration which takes advantage of the intricate polarization dependencies in the interacting optically driven polariton system. Besides detailed numerical simulations of the coupled light-field exciton dynamics, in the present paper we focus on the derivation of a simplified population competition model giving detailed insight into the underlying mechanisms from a nonlinear dynamical systems perspective. We show that such a model takes the form of a generalized Lotka-Volterra system for two competing populations explicitly including a source term that enables external control. We present a comprehensive analysis both of the existence and stability of stationary states in the parameter space spanned by spatial anisotropy and external control strength. We also construct phase boundaries in  non-trivial regions and characterize emerging bifurcations. The population competition model reproduces all key features of the switching observed in full numerical simulations of the rather complex semiconductor system and at the same time is simple enough for a fully analytical understanding.
\end{abstract}

\maketitle

\section{\label{sec:intro}Introduction}

The demand for integrated optoelectronic devices in optical communication networks has resulted in an increase of research activities targeted at functional all-optical components. For example, a wide range of different approaches has been proposed to realize efficient all-optical switches exploiting the nonlinear optical properties of different material systems, including organic photonic crystals \cite{Hu2008}, rubidium atomic-vapor cells \cite{Dawes672}, or GaAs semiconductor microcavities as in Refs. \cite{ballarini2013all,amo2010exciton,sanvitto2016road} and Refs. \cite{Kheradmand2008,Schumacher2009,PhysRevB.87.205307}. The latter utilize the optical control of transverse optical patterns to achieve transistor-like switching performance \cite{Dawes672,Lewandowski17}. Spontaneous formation of spatially extended patterns has been intensively studied during the past decades \cite{cross1993} with applications to different areas of science including formation of sand ripples and desert sand dunes \cite{bowmannewell98}, animal coat patterns such as zebra stripes \cite{Meinhardt1982,murray03}, or geographic patterns in parasitic insect populations \cite{hassel}. With applications to optical switching, however, the quest for efficient external control of these patterns naturally arises, but in many cases has not been explored in detail. In the present work we investigate a specific example of all-optical switching of polariton patterns in a semiconductor quantum-well microcavity system. In contrast to our previous work \cite{Lewandowski17} we give a detailed analysis of the optical switching dynamics from a nonlinear dynamical systems perspective. To this end we derive a simplified mode competition model that governs the essentials of the system dynamics but is simple enough to fully characterize the possible stationary solutions and phase-space singularities in the relevant parameter space. This simplified model has the very general mathematical form of a generalized Lotka-Volterra system, with the addition of an inhomogeneity for external control. In the main part of the paper (cf. Sec. \ref{sec:PC_model}), we present a complete steady state analysis of this nonlinear system. The solution space we obtain in dependence of spatial anisotropy and external control strength is of very general nature and may be similarly realized in other systems where external control of population competition is studied such as chemical reactions or in the life sciences.
 
\begin{figure}[t]
		\centering
		\includegraphics[width=0.48 \textwidth]{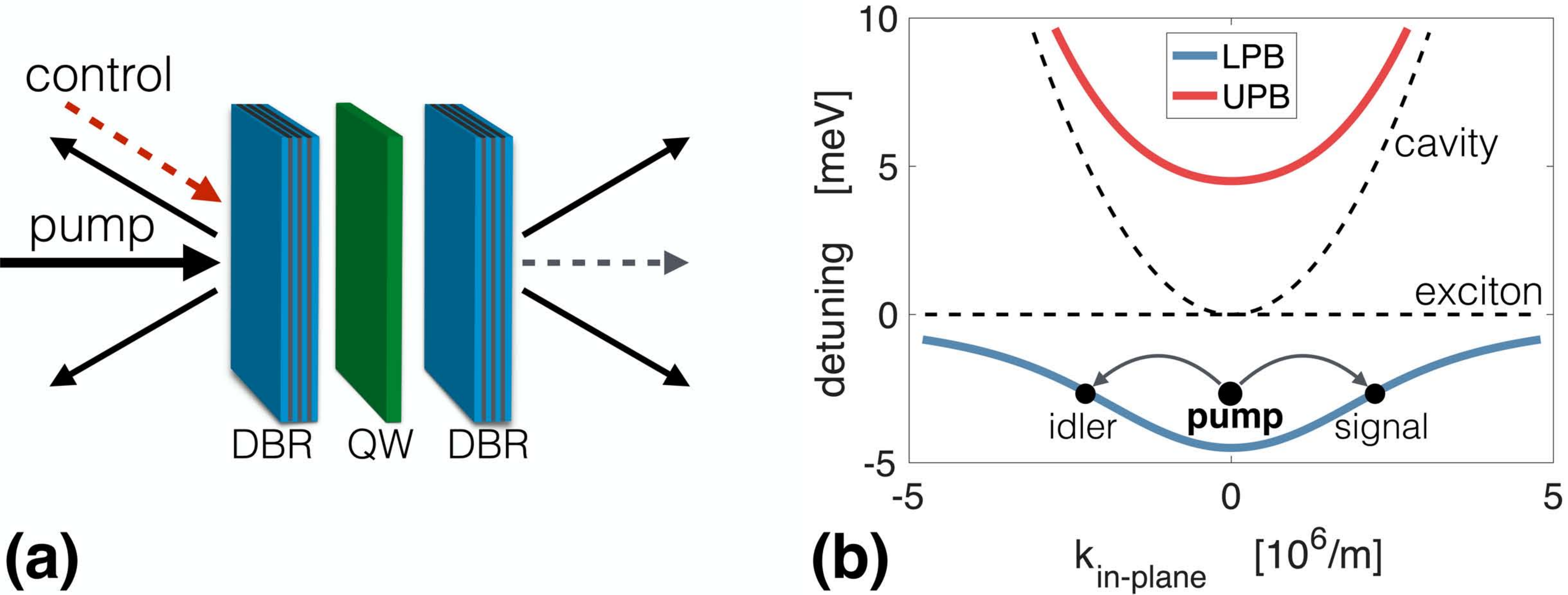}
		\caption{(Color online) (a) Sketch of a planar quantum-well semiconductor microcavity with continuous-wave pump at normal incidence (on-axis) and optional off-axis control beam. (b) Corresponding polariton dispersion relation in the normal-mode splitting regime. The dominant four-wave mixing process is indicated, driving the pairwise scattering of pump-induced polaritons onto the lower polariton branch with opposite in-plane momenta for signal and idler modes.}
		\label{fig:microcavity}
\end{figure}

As our specific example, here we study planar semiconductor microcavities with a strong coupling between the cavity field and the exciton polarization that gives rise to the formation of exciton polaritons \cite{Kavokin}. These quasiparticles consist of a photonic and an excitonic part and are characterized by a normal-mode splitting in the dispersion relation, on the lower branch leading to long coherence times and strong nonlinear interactions. The latter are driven by four-wave mixing processes of coherent polariton fields which can also be interpreted as polariton-polariton scattering, mediated through exciton-exciton scattering. For certain excitation conditions, small spatially varying density fluctuations can experience huge growth in particular modes due to the intrinsic feedback mechanisms driven by four-wave mixing. This causes spatially homogeneous density distributions to become unstable such that the system's symmetry is spontaneously broken. This results in the formation of stationary patterns, directly observable in the far field emission from the microcavity. Figure~\ref{fig:microcavity}(a) shows the excitation geometry used with the pump at normal incidence (zero in-plane momentum) and finite off-axis ($k{\neq}0$) signals due to amplified fluctuations. Figure~\ref{fig:microcavity}(b) shows the normal-mode splitting of the dispersion relation into a lower (LPB) and upper polariton branch (UPB) alongside the bare exciton and cavity dispersions. Phase-matching conditions determine the efficiencies of the different scattering processes, and therefore determine the possible pattern geometry. For scalar (only one circular polarization state) polariton fields hexagon patterns are favored \cite{ardizzone,egorov2014motion,saito2013order}. Extending the model with polarization dependence a complex interplay of the TE/TM cavity-mode splitting in the linear regime and the spin-dependent exciton-exciton interactions in the nonlinear regime arises. The resulting polarization-induced spatial anisotropy \cite{PhysRevB.94.045308} determines the possible unstable modes and stable patterns formed. For linearly polarized pump excitation slightly above instability threshold, cross-linearly polarized resonant modes parallel and perpendicular to the pump's polarization plane constitute the basic instabilities \cite{PhysRevB.77.073302}, resulting in two-spot or four-spot patterns. Making use of this spatially anisotropic polariton-polariton scattering a reversible optical switching for orthogonal two-spot patterns can be realized that is triggered by a weak external control beam \cite{Lewandowski17}. The advance of taking the polarization dependence into account, compared to scalar switching schemes \cite{Kheradmand2008,Schumacher2009,PhysRevB.87.205307}, is the explicit breaking of the rotational symmetry, offering to control the orientation of the initial pattern via a linearly polarized pump and to provide an automatic back-switching mechanism. It also makes background-free detection possible, since the pattern switching takes place in the cross-polarized channel.

\section{\label{sec:polaritons}Orthogonal Switching of Two-Spot Patterns}
\begin{figure}[t]
		\centering
		\includegraphics[width=0.48 \textwidth]{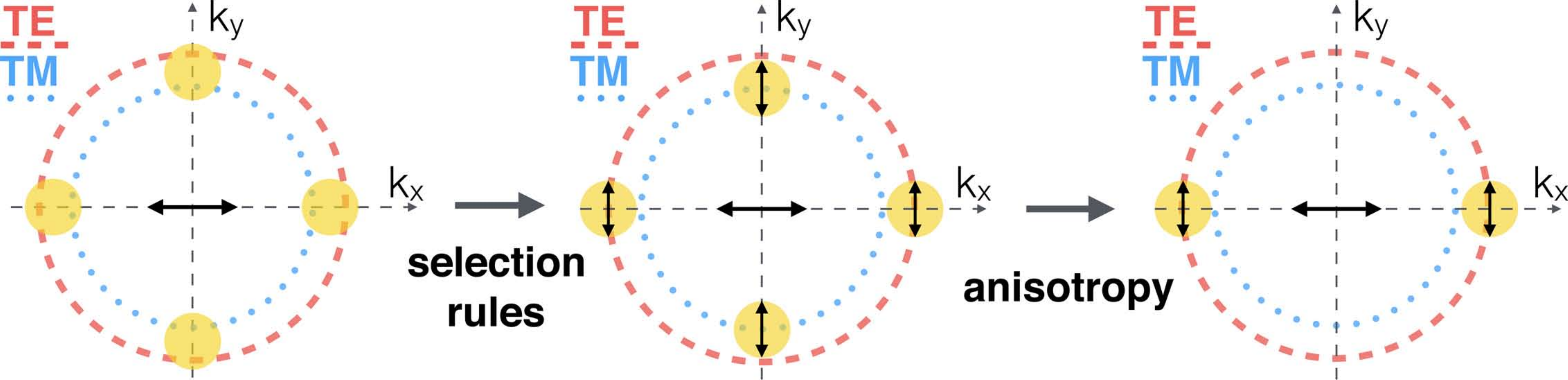}
		\caption{(Color online) Schematic illustration of the dominant contributions to the in-plane scattering of polaritons onto the elastic circles defined by the TE and TM modes of the lower polariton branch. The pump indicated in the center is $x$-polarized and slightly above off-axis instability threshold. Parallel and perpendicular to the pump polarization direction, the scattering can occur in either TE or TM mode (left panel). For pumping spectrally well below the exciton resonance, the scattering with polarization orthogonal to the pump's is preferred (center panel). With spatial anisotropy (which is partly induced by the linear pump polarization) scattering along the pump's polarization direction onto the TE mode can dominate (right panel).}
		\label{fig:orth_switching}
\end{figure}

A detailed numerical investigation of the present system for the following excitation conditions was already discussed in Ref.~\cite{Lewandowski17}. The system is excited (driven) by an $x$-linearly polarized continuous wave pump at normal incidence with flat-top Gaussian shape in the QW plane and an intensity slightly above the off-axis instability threshold. In this case spontaneous breaking of spatial and polarization symmetry is observed. Signals are formed by resonant polariton scattering onto the elastic circle defined by the polariton dispersion (cf. Fig. \ref{fig:microcavity}). As can also be understood based on a linear stability analysis \cite{schumacher2007influence,PhysRevB.77.073302}, the scattering occurs predominantly in four spatial directions oriented orthogonal and parallel to the pump polarization plane, respectively. In general, in any spatial direction signals can form either in the TE or in the TM mode as illustrated in Fig. \ref{fig:orth_switching}. However, for the pump polaritons scattered to finite $k$ through Coulomb interaction, due to the underlying spin-dependent exciton-exciton interactions, the scattering probability is higher for a polarization state orthogonal to the pump polarization state \cite{schumacher2007influence}. Therefore, in the spatial direction parallel to the pump polarization state, the scattered signals preferentially form in the TE mode, and in the TM mode for scattering orthogonal to the  pump polarization state (cf. Fig. \ref{fig:orth_switching}). Out of these four signals, a stationary two-spot pattern (only one mode pair with opposite in-plane momenta) can be prepared by introducing some anisotropy favouring one direction over the other. Alongside a slight polarization induced spatial anisotropy that is due to a slightly higher density of states for the TE modes \cite{PhysRevB.77.073302}, an additional anisotropy can be introduced by tilting the pump beam slightly away from normal incidence (cf. Fig. \ref{fig:orth_switching}). In the nonlinear system studied, at sufficiently high densities of the favored two-spot pattern, cross-saturation processes will lead to the extinction of the signals in the other orthogonal direction. This allows us to single out a two-spot pattern $\mathrm{T}_1$ which will serve as the initial state in the switching process. Here, we start with the two-spot pattern oriented in the $x$-direction (direction of intrinsic anisotropy). This selection process is schematically visualized in Fig.~\ref{fig:orth_switching}. 

\begin{figure}[t]
		\centering
		\includegraphics[width=0.48 \textwidth]{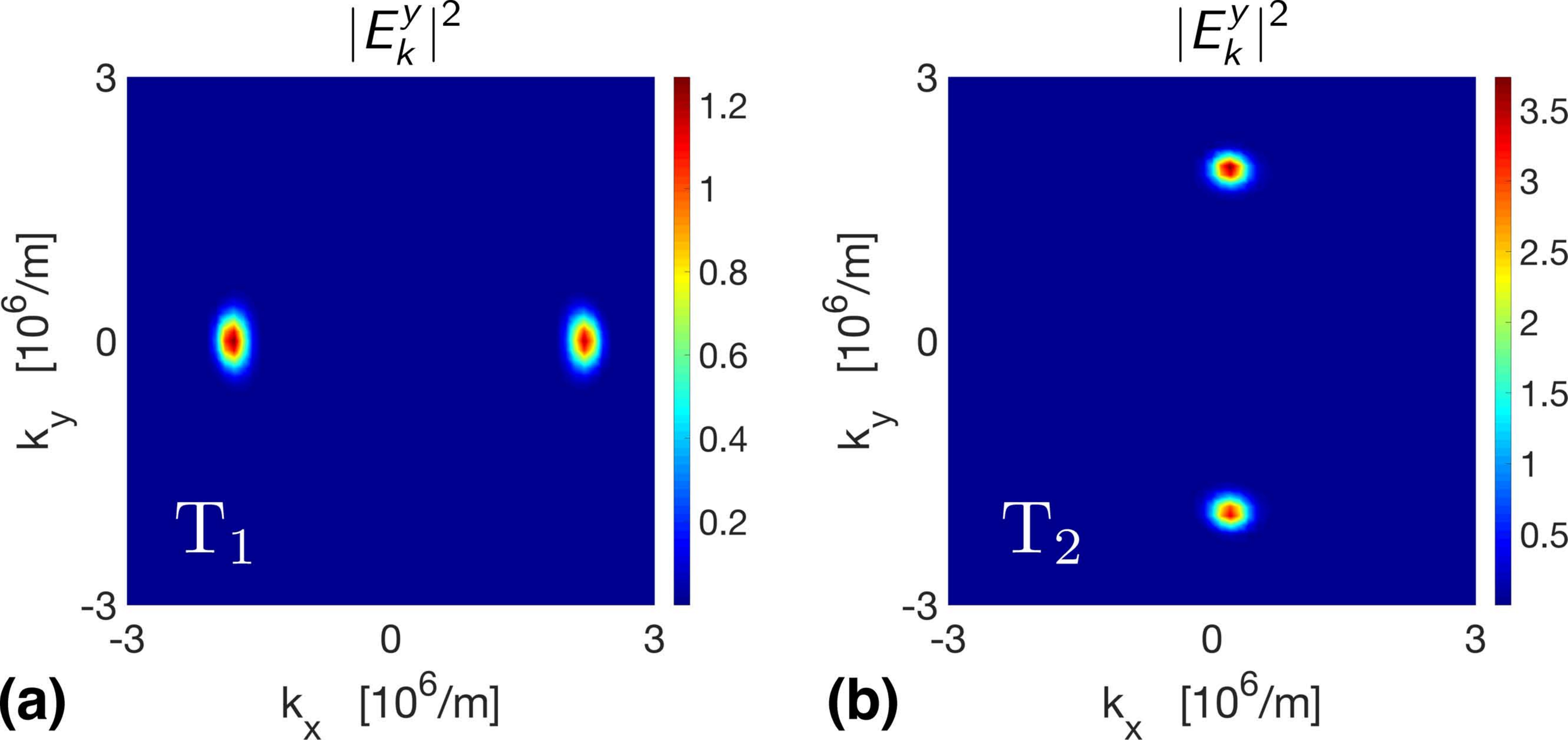}
		\caption{(Color online) (a) Stationary $\mathrm{T}_1$ and (b) $\mathrm{T}_2$ state of the polariton pattern switch as obtained from the numerical solution of Eqs. \eqref{eq:full model} for $x$-linearly polarized cw pump at $k\approx 0$. Shown is the stationary $y$-polarization component of $|E|^2$ in $k$-space in arbitrary units (a) before and (b) after switching on a continuous wave control beam. Switching off the control beam results in reversal of the switching process such that after sufficiently long times the system returns to the stationary $\mathrm{T}_1$ pattern.}
		\label{fig:sim_result}
\end{figure}

If now a weak (compared to the pump intensity) $y$-polarized control beam with an in-plane momentum on the elastic circle and spatially orthogonal to the initial pattern is applied, stimulated scattering leads to population revival of the corresponding two-spot pattern. Due to the cross-saturation effect, the initial pattern is destabilized and finally switched off completely while the orthogonal pattern reaches a steady state $\mathrm{T}_2$ (cf. Fig. \ref{fig:sim_result}). After the control beam has been switched off again, the anisotropy leads to re-emergence of the initial $\mathrm{T}_1$ pattern while the $\mathrm{T}_2$ state vanishes again. If the control beam is too weak so that complete switching may not be possible, the system will remain in a stationary four-spot pattern state $\mathrm{F}$. Based on this scheme, in Ref.~\cite{Lewandowski17} transistor-like reversible switching was demonstrated including a systematic study of switching times, minimum control power needed, and achievable gain. This was done by numerical simulations of the nonlinear set of equations of motion governing the coherent coupled light field and exciton dynamics in the microcavity system in the two-dimensional QW plane in real space, 
\begin{equation}\label{eq:full model}
\begin{aligned}
\mathrm{i}\hbar\partial_{t}E^{\pm}=&(H_c-\mathrm{i}\gamma_c) E^{\pm}+H^{\pm} E^{\mp} - \Omega p^{\pm} + E^{\pm}_{\mathrm{pump}} \\
\mathrm{i}\hbar\partial_{t}p^{\pm}=&(\epsilon_0^e-\mathrm{i}\gamma_e)p^{\pm} - \Omega (1-\alpha_{\mathrm{psf}}|p^{\pm}|^2)E^{\pm} \\
&+T^{++}|p^{\pm}|^2p^{\pm}+T^{+-}|p^{\mp}|^2p^{\pm}.
\end{aligned}
\end{equation}
Here, the index $\pm$ denotes the different components in the circular polarization basis, $H_c{=}\hbar\omega_0{-}\tfrac{\hbar^2}{4}(\tfrac{1}{m_{\mathrm{TM}}}{+}\tfrac{1}{m_{\mathrm{TE}}})(\partial_x^2{+}\partial_y^2)$ is the cavity Hamiltonian and $\epsilon_0^e$ the flat exciton dispersion. $\gamma_c$ and $\gamma_e$ represent the photon and exciton loss rates, $\Omega$ describes the photon-exciton coupling, and $H^{\pm}{=}{-}\tfrac{\hbar^2}{4}(\tfrac{1}{m_{\mathrm{TM}}}{-}\tfrac{1}{m_{\mathrm{TE}}})(\partial_x {\mp} \mathrm{i} \partial_y)^2$ couples the two polarization components due to TE/TM-splitting. The cubic nonlinearities consist of a phase-space filling term $\alpha_{\mathrm{psf}}$, repulsive interaction $T^{++}$ for excitons with parallel spins, and attractive interaction $T^{+-}$ for excitons with opposite spins.

Based on Eqs. \eqref{eq:full model} with an $x$-linearly polarized continuous wave pump beam at $k\approx 0$ we perform a numerical simulation to demonstrate the basic switching process in the two-dimensional plane of in-plane momenta. Figure~\ref{fig:sim_result} shows the photonic component, $|E|^2$, in $k$-space without control beam such that the stationary pattern $\mathrm{T}_1$ forms in panel (a) and with the control beam on, switching to the stationary pattern $\mathrm{T}_2$ in panel (b). The control beam has the same frequency as the pump beam (cl. Fig. \ref{fig:microcavity}). System parameters used and details of the calculations are given in Appendix \ref{app:sec:Numerical Details}. Upon switching off the control beam the switching action is reversed and the pattern returns to its original state $T_1$. Switching times in the range of ${\sim}100~\mathrm{ps}$ are achievable \cite{Lewandowski17}.

The focus of the present paper is not on the full numerical simulations of Eqs. \eqref{eq:full model} and resulting switching performance. Rather, starting from Eqs. \eqref{eq:full model} we will derive a simplified population competition model for selected modes in $k$-space (details of the derivation are given in Appendix B) to provide further insight into the underlying phase space singularities that dictate the global behavior and solution space of the nonlinear dynamical system studied. In a similar fashion this approach was previously applied to the switching between subsets of a hexagonal pattern \cite{PCmodel}. We will systematically analyze the existence and stability properties of possible steady states in dependence of the strength of the different involved physical processes for a typical orthogonal switching setup comparable to the one introduced above. To this end, we will construct phase boundaries in representative regions of parameter space and characterize relevant bifurcations. This will lead to a general understanding of crucial parameter dependencies such as the ratio between the control beam strength and the anisotropy, and also of the coexistence of solutions in certain parameter regions and hysteresis behavior. Based on the simplified model, we will also be able to show that the polariton dynamics and optical switching phenomena discussed in the present paper can mathematically be understood based on an generalized Lotka-Volterra model including an external control parameter.  

\section{\label{sec:PC_model}Population Competition Model}
\begin{figure}[t]
		\centering
		\includegraphics[width=0.48 \textwidth]{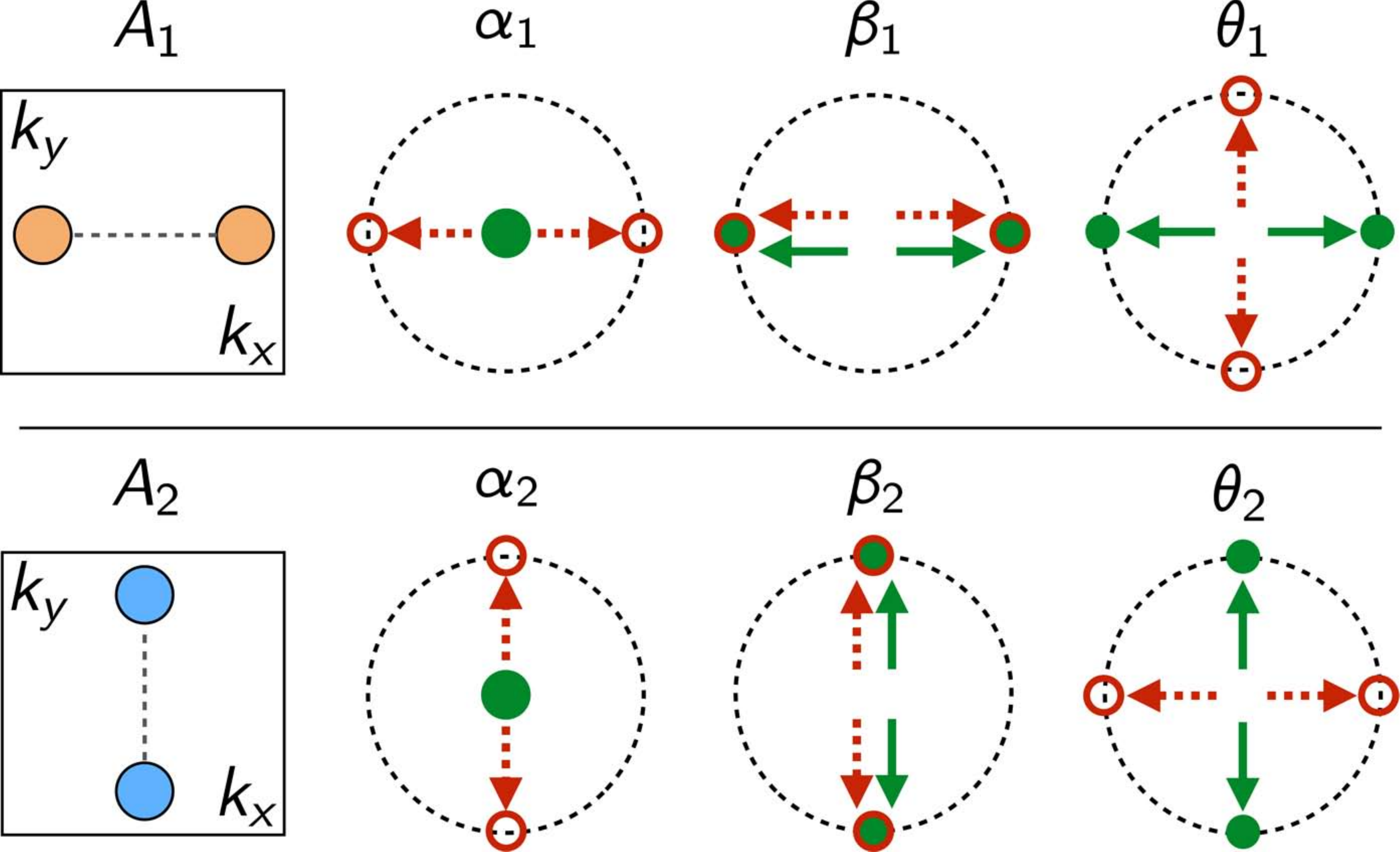}
		\caption{(Color online) Illustration of the different polariton-polariton scattering processes in $k$-space for the two elementary mode-pairs $A_1$ and $A_2$ in Eq.~(2). The dashed circle represents ring in $k$-space on which efficient scattering is possible. \textit{Green solid} (\textit{red open}) circles mark the incoming (outgoing) modes. \textit{Solid} (\textit{dashed}) arrows mark their corresponding momenta. The $\alpha_i$ describe the growth process of the two mode pairs (basic instability leading to pattern formation) and the $\beta_i$ and $\theta_i$ represent the self- and cross-saturation process.}
		\label{fig:processes}
\end{figure}

The simplified population competition model discussed in the remainder of the present paper can be derived from the set of equations of motion in Eq.~(1) as detailed in Appendix \ref{app:sec:derivation}. It reads
\begin{equation}\label{eq:PC model}
\begin{aligned}
\dot{A}_1 &=\alpha_1 A_1 - \beta_1 A_1^3 - \theta_1 A_2^2 A_1 \\
\dot{A}_2 &=\alpha_2 A_2 - \beta_2 A_2^3 - \theta_2 A_1^2 A_2+S.
\end{aligned}
\end{equation}
Here $A_1$ and $A_2$ are real-valued amplitudes of the two elementary states of the system, i.e., the two orthogonal polariton mode pairs in $k$-space. In case of stationary solutions the phases in the full model \eqref{eq:full model} become “locked”, allowing us to remove them as dynamical variables from the equations which is suitable for the following steady state analysis. The six dimensionless real-valued positive parameters, $\alpha_i$, $\beta_i$, $\theta_i$ are directly related to the main (up to third order) polariton-polariton scattering processes of the original system as illustrated in Fig.~\ref{fig:processes}. They can be calculated from the physical parameters of the full model (cf. Appendix \ref{app:sec:parameters}). These parameters are intrinsically different for $A_1$ and $A_2$ due to the polarization dependence and anisotropy. The linear process representing growth of the resonant modes is described by $\alpha_i$. Saturation processes are represented by the cubic terms which can be divided into self-saturation ($\beta_i$) and cross-saturation ($\theta_i$). The external control is described by the inhomogeneity $S$.
If we rewrite Eqs.~(\ref{eq:PC model}) as $(\dot{A}_1, \dot{A}_2)^{\mathrm{T}} \equiv (f_1,f_2)^{\mathrm{T}}= \mathbf{f}$, steady states are characterized by $\mathbf{f}{=}\mathbf{0}$. Four qualitatively different stationary solutions are possible: i) two-spot pattern $\mathrm{T}_1$ with $A_1{\neq}0$ and $A_2{=}0$, ii) two-spot pattern $\mathrm{T}_2$ with $A_1{=}0$ and $A_2{\neq}0$,  iii) four-spot pattern $\mathrm{F}$ with $A_1{\neq}0$ and $A_2{\neq}0$, and iv) the trivial solution with $A_1{=}0$ and $A_2{=}0$. Here we are only interested in physical solutions $A_{i}\geq 0$, and therefore the state space is $\mathbb{R}^2_{\geq 0}$, i.e. the first quadrant of the $(A_1,A_2)$-plane. The linear term ($\alpha$) leads to exponential growth of the corresponding mode pair and the self-saturation ($\beta$) has a stabilizing effect. For these two processes, the equations (\ref{eq:PC model}) are automatically decoupled and may be solved separately, resulting in a stable four-spot pattern. However, the cross-saturation terms ($\theta$) couple the two modes and tend to suppress a particular mode pair, favoring the other mode pair, resulting in a two-spot pattern. Additionally, the external control acts as a source term for $A_2$. The PC model thus describes the dynamical competition between the three types of possible stationary patterns: $\mathrm{T}_1$, $\mathrm{T}_2$ and F. 
A phase in the PC model is defined as a set consisting of the number of steady states and their stability properties. These sets, i.e. the phase, are functions of the seven parameters and a phase boundary in parameter space indicates the change in the number of steady states and/or their stability. This can only happen at points where at least one eigenvalue of the corresponding Jacobian matrix $J{\equiv}\left( \partial_{A_j} f_i \right)$ is zero \cite{wiggins2006introduction}, which is equivalent to the condition $\det{J}{=}0$. We thus find phase boundaries in parameter space at points satisfying $\lbrace \mathbf{f},\det{J} \rbrace{=}\mathbf{0}$.

\subsection{\label{subsec:hom_case}Homogeneous Case $S=0$}
The homogeneous case of Eq.~\eqref{eq:PC model} ($S=0$) has the form of a generalized Lotka-Volterra (GLV) model \cite{brenig1988complete} with cubic nonlinearities. The transformation $A_i\rightarrow A_i^2\equiv\tilde{A}_i$ yields the usual Lotka-Volterra (LV) equations \citep{hofbauer1988theory} with quadratic nonlinearities while conserving the $(A_1,A_2)$ phase space structure in the positive quadrant \cite{hernandez1997lotka}. Hence, for $S=0$ the phase portraits of \eqref{eq:PC model} are topologically equivalent to those of the following well-known system
\begin{equation}\label{eq:LV_equation}
\partial_t \tilde{A}_i=\tilde{A}_i \left( r_i  - \sum_{j=1}^{2} c_{ij} \tilde{A}_j \right)
\end{equation}
with $i=1,2$, growth rate vector $\mathbf{r}=2(\alpha_1,\alpha_2)$, and community matrix 
\begin{equation}
C=2\begin{pmatrix}
\beta_1 & \theta_1 \\ \theta_2 & \beta_2
\end{pmatrix},
\end{equation}
which describes self- and cross-interaction. This LV model \eqref{eq:LV_equation} for interspecific competition has been studied in many different contexts, e.g. ecology \cite{wangersky1978lotka}, chemistry \cite{Hering1990}, economics \cite{goodwin1982growth}, physics \cite{lamb1964theory} and it was shown \cite{roughgarden1979theory,hofbauer1988theory} that its dynamical behavior is limited to three cases depending on the parameters:
i) Coexistence regime for $\mathrm{sgn}(\det{C}){=}{+}1$ (larger self-saturation), ii) bistability regime for $\mathrm{sgn}(\det{C}){=}{-}1$ (larger cross-saturation), iii) dominance regime for sufficiently large/small growth rate ratios. In this regime long-time dynamics are independent of the interaction parameters and always result in extinction of one population. Although the LV system, Eq.~\eqref{eq:LV_equation}, is well-known, we present a short discussion of the homogeneous case of the PC model, Eq.~\eqref{eq:PC model}, and point out the importance for application to the polariton switching dynamics. Since for $S=0$ the system is solvable analytically we obtain explicit expressions for all steady states and phase boundaries: i) $\mathrm{T}_{\mathrm{1}}$: $A_1{=}\sqrt{\tfrac{\alpha_1}{\beta_1}}$, $A_2{=}0$, stable for $\tfrac{\alpha_1}{\alpha_2}>\tfrac{\beta_1}{\theta_2}$, ii) $\mathrm{T}_{\mathrm{2}}$: $A_1{=}0$, $A_2{=}\sqrt{\tfrac{\alpha_2}{\beta_2}}$, stable for $\tfrac{\alpha_1}{\alpha_2}<\tfrac{\theta_1}{\beta_2}$, iii) $\mathrm{F}$: $A_1{=}\sqrt{\tfrac{\alpha_2 \theta_1 - \alpha_1 \beta_2}{\theta_1 \theta_2 - \beta_1 \beta_2}}$, $A_2{=}\sqrt{\tfrac{\alpha_1 \theta_2 - \alpha_2 \beta_1}{\theta_1 \theta_2 - \beta_1 \beta_2}}$, only exists for $\tfrac{\beta_1}{\theta_2}{\lessgtr}\tfrac{\alpha_1}{\alpha_2}{\lessgtr} \tfrac{\theta_1}{\beta_2}$ and is stable for $\theta_1 \theta_2 {<} \beta_1 \beta_2$, iv) trivial solution: $A_1{=}A_2{=}0$.
We already see that the four-spot solution does not exist in the entire $S{=}0$ plane in contrast to the two-spot pattern. The trivial solution also exists everywhere but is always unstable since the eigenvalues of its Jacobian $\alpha_1,\alpha_2$ are positive. This solution will not be listed hereinafter. For a systematic discussion we introduce a variable anisotropy parameter $\delta\alpha$ in the first equation of \eqref{eq:PC model}, $\alpha_1\rightarrow\alpha_1+\delta\alpha$, and set $\alpha_1{=}\alpha_2{=}1$. It favors $A_{1/2}$ for $\delta\alpha\gtrless 0$, respectively. 
\begin{figure}[t]
		\centering
		\includegraphics[width=0.44 \textwidth]{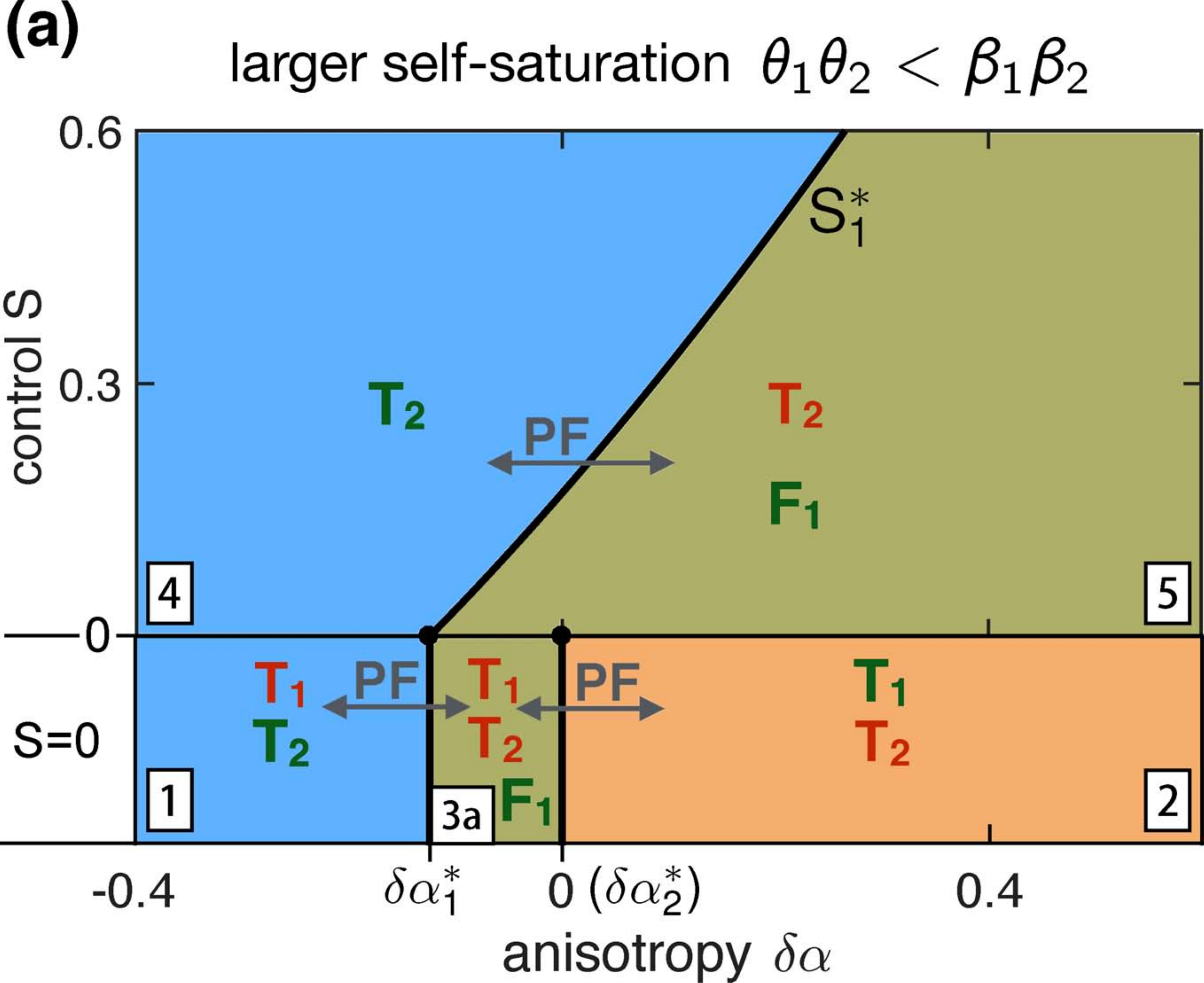}
		\includegraphics[width=0.44 \textwidth]{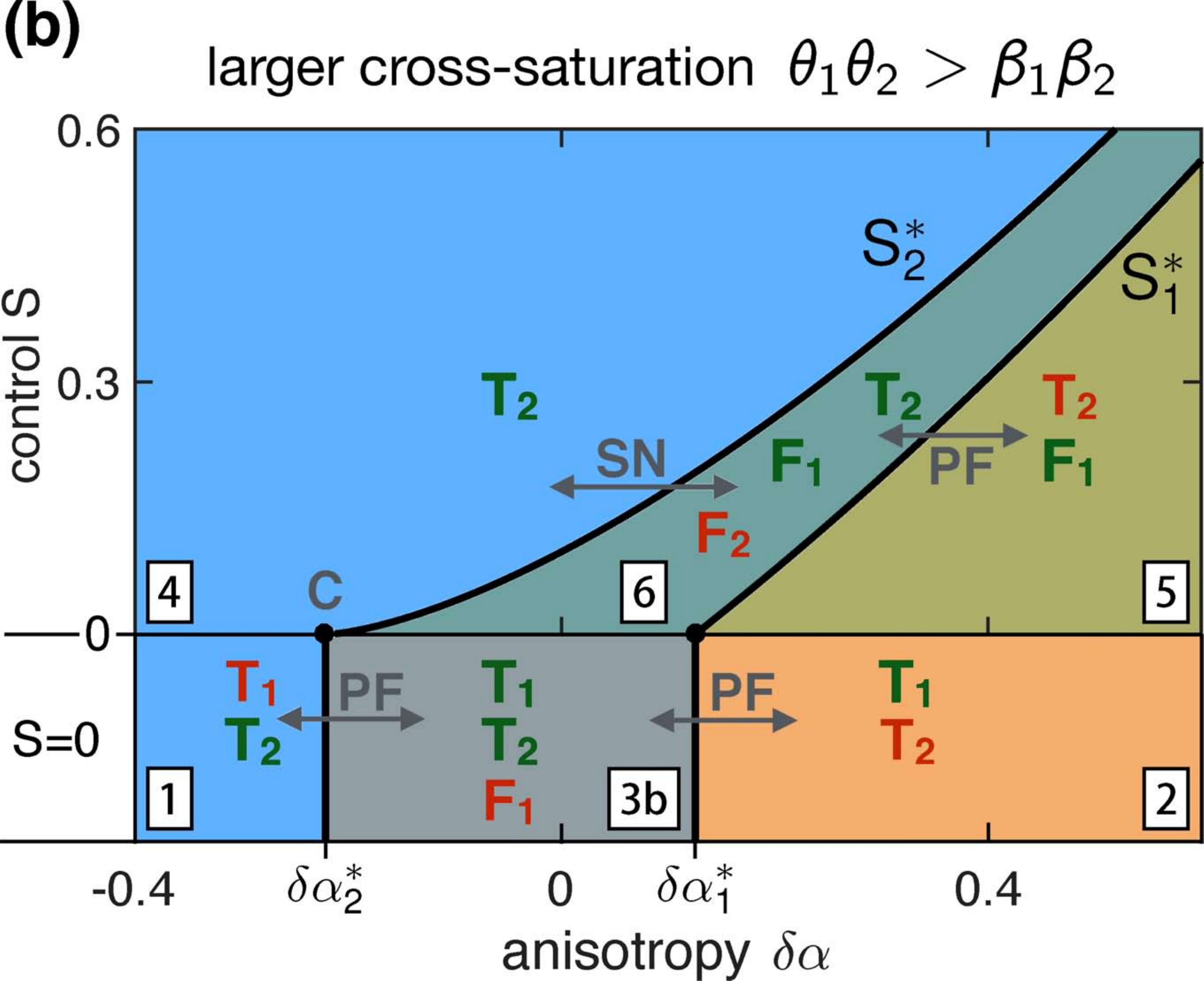}
		\caption{(Color online) Phase boundaries in $(\delta\alpha,S)$ parameter space showing regions of different stable and unstable steady states in dependence of the anisotropy and external control. (a) Larger self-saturation $\beta_1 \beta_2{>}\theta_1 \theta_2$. (b) Larger cross-saturation $\beta_1 \beta_2{<}\theta_1 \theta_2$. \textit{Green} (\textit{dark gray}) / \textit{red} (\textit{light gray}) letters mark stable / unstable steady states. Lines $S_1^*$ and $S_2^*$ show the phase boundaries \eqref{eq:boundary1} and \eqref{eq:boundary2}.}
		\label{fig:phase_portraits}
\end{figure}
The resulting phase boundaries are shown in Fig. \ref{fig:phase_portraits} where the homogeneous case ($S{=}0$) is included in the extended region at the bottom of each plot. It shows the structure of the usual LV model.
\textit{Green} (\textit{dark gray}) / \textit{red} (\textit{light gray}) letters mark stable / unstable steady states. For large anisotropy $|\delta\alpha|$ only stable and unstable two-spot patterns are possible. This dominance regime does no longer depend on the interaction parameters $\beta_i$ and $\theta_i$. The middle region on the other hand is divided into two cases: A stable four-spot pattern for larger self-saturation ($\beta_1 \beta_2{>}\theta_1 \theta_2$, coexistence, upper panel) and two simultaneously stable two-spot patterns for larger cross-saturation ($\beta_1 \beta_2{<}\theta_1 \theta_2$, bistability, lower panel). 

\begin{figure}[t]
		\centering
		\includegraphics[width=0.48 \textwidth]{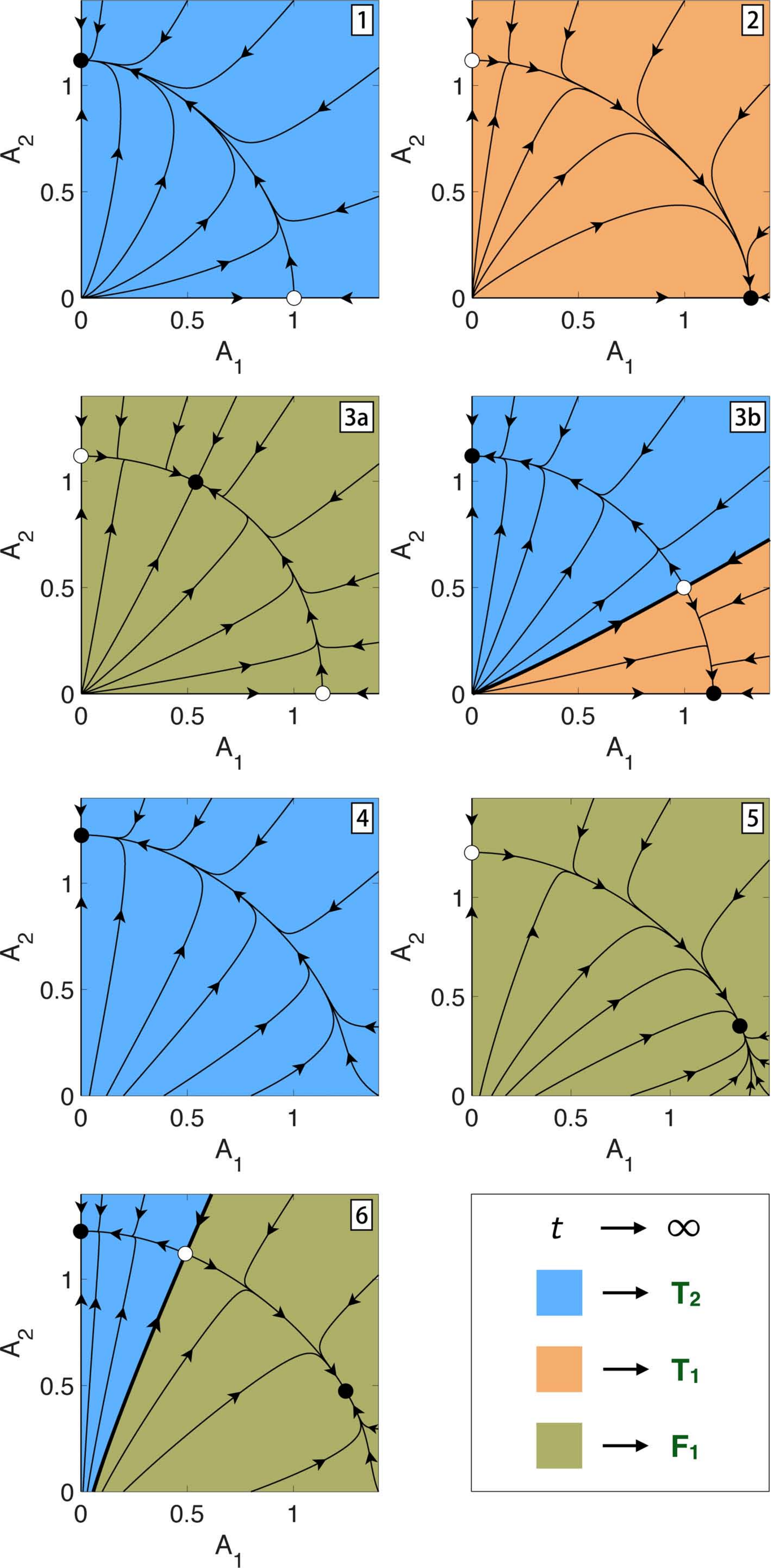}
		\caption{(Color online) Phase plane flow for cases corresponding to Fig.~\ref{fig:phase_portraits}. Black (white) dots mark stable (unstable) steady states. Black lines show representative orbits. The thick black line in 3b and 6 is the stable manifold of the saddle point which separates two basins of attraction.}
		\label{fig:flow}
\end{figure}

The first four cases in Fig.~\ref{fig:flow} show the corresponding flow given by representative trajectories in state space for the homogeneous case. The system's dynamics are unambiguous in cases 1, 2, and 3a where only one attractor exists which determines the long time behavior. However, in case 3b two attractors exist and the dynamics now depend on the system's history leading to a hysteresis effect. The two basins of attraction are marked with different colors \textit{blue} (\textit{dark gray}) and \textit{orange} (\textit{light gray}). They are separated by the stable manifold of the saddle point F (often called \textit{separatrix}), defined by the set of points $(A_1,A_2)$ which satisfy $(A_1,A_2)\rightarrow\mathrm{F}$ for $t\rightarrow\infty$. In contrast to this, the line connecting all three steady states is the unstable manifold of F, consisting of points which satisfy $(A_1,A_2)\rightarrow\mathrm{F}$ for $t\rightarrow -\infty$. Any initial point will stay in its region (\textit{blue} or \textit{orange}) and end up at the corresponding attractor ($\mathrm{T}_1$ or $\mathrm{T}_2$). The system thus shows hysteresis behavior which might prevent complete back-switching, and therefore should be avoided for switching purposes, for example by increasing the anisotropy.

The Hartman Grobman theorem \cite{wiggins2006introduction} ensures that the behavior near hyperbolic equilibrium points is completely determined by its linearization. This no longer holds for non-hyperbolic fixed points which are characterized by the existence of at least one zero real-part eigenvalue of the corresponding Jacobian. Therefore, the behavior at the phase boundaries in parameter space can not be analyzed via linear stability analysis. Instead one can use center manifold theory and normal forms \cite{wiggins2006introduction} in order to determine the occurring bifurcations. They are transcritical for the usual LV model but different in the case of cubic nonlinearities due to the additional $\mathbb{Z}_2 \times \mathbb{Z}_2$ symmetry. For both two-spot solutions symmetric pitchfork (PF) bifurcations occur at the phase boundaries when the anisotropy parameter is changed. They are supercritical for $\beta_1 \beta_2{>}\theta_1 \theta_2$ (larger self-saturation) leading to the coexistence regime with a stable F pattern and subcritical for $\beta_1 \beta_2{<}\theta_1 \theta_2$ (larger cross-saturation) leading to the bistability regime with stable $\mathrm{T}_1$ and $\mathrm{T}_2$ pattern. Pitchfork bifurcations are typical for dynamical systems with inversion symmetry, here $A_i\rightarrow {-}A_i$. Since we are only interested in positive solutions in the first quadrant of the phase plane, we either observe a transition from a stable T solution to an unstable T and a stable F solution (supercritical) or the same transition with reversed stability (subcritical).

In conclusion, all phase portraits of the homogeneous case can be completely reduced to the results of the well-known Lotka-Volterra model for two competitive species. Due to the additional inversion symmetry we obtain pitchfork bifurcations at the phase boundaries. The homogeneous case is crucial for the polariton switch because it describes the initial pattern formation and back-switching process in absence of the control beam. Both can only work reliably in regions where only the $\mathrm{T}_1$ pattern exists as a single attractor. Therefore, a sufficient minimum anisotropy in favor of $A_1$ is needed. We have also seen that if the cross-saturation dominates over the self-saturation, the coexistence of $A_1$ and $A_2$ is destabilized resulting in the extinction of $A_i$ and survival of $A_{i+1}$. A strong interspecific competition thus prevents the coexistence. Otherwise, if the self-saturation is stronger than the cross-saturation, the dominating intraspecific competition promotes coexistence. The parameters used for the full numerical simulations in Sec. \ref{sec:polaritons} lead to the larger self-saturation case and sufficient anisotropy (cf. Appendix \ref{app:sec:parameters}), providing reliable initial pattern formation and back-switching, since the F pattern is unstable.

\subsection{\label{subsec:inhom_case}Inhomogeneous Case $S>0$}
The inhomogeneous ($S{>}0$) case of the PC model \eqref{eq:PC model} describes the actual switching process induced by an external control beam. This term can also be motivated for other systems where the GLV model is commonly used, e.g. to include constant migration/harvesting in the description of ecological systems or constant influx in chemical reactions. Therefore, it can be interpreted as an extension of the generalized Lotka-Volterra dynamics and the following analysis is of very general interest but has not been investigated before. The influence of constant terms in the usual LV model with quadratic nonlinearities was investigated in Ref. \cite{Saputra2010} from a purely mathematical perspective, but in general inhomogeneous population competition models did not receive much attention in the past. Here, we analyze and apply the GLV model with a constant inhomogeneity to the case of the orthogonal switching of two-spot polariton patterns. 

Also including inhomogeneity, all steady states and phase boundaries can still be determined analytically due to the system's simplicity. However, we note that in general solving multivariate polynomial equation systems is a difficult task which can be simplified using algebraic methods (e.g. Gr\"obner Basis \cite{buchberger1,buchberger2}), as was shown in Ref.~\cite{PCmodel}. The non-zero source term $S$ in \eqref{eq:PC model} breaks the inversion symmetry for $A_2$ and prevents a $\mathrm{T}_1$ solution. This leaves us with only two competing patterns (called $\mathrm{T}_2$ \& $\mathrm{F}_2$) this time around. However, as we will see below it includes the possibility of an additional qualitatively different four-spot pattern solution $\mathrm{F}_1$. The control parameter $S$ acts as a constant source term for the $A_2$ mode pair, leading to linear growth. Supporting $A_2$ means simultaneously suppressing $A_1$ due to the cross-saturation effect, leading to a more asymmetric $\mathrm{F}_2$ solution in favor of $A_2$. A second four-spot pattern, $\mathrm{F}_1$, which is in favor of $A_1$ replaces $\mathrm{T}_1$. The phase boundaries are given explicitly by
\begin{equation}\label{eq:boundary1}
S^*_1(\delta\alpha)= \beta_2 \sqrt{\frac{1+\delta\alpha}{\theta_1}}^{~3}- \sqrt{\frac{1+\delta\alpha}{\theta_1}},
\end{equation}
and additionally, for the case of larger cross-saturation ($\beta_1 \beta_2{<}\theta_1 \theta_2$)
\begin{equation}\label{eq:boundary2}
S^*_2(\delta\alpha)=6\sqrt{\frac{3\left[ (1+\delta\alpha)\theta_2-\beta_1 \right]^{3}}{\beta_1^2 (\theta_1\theta_2-\beta_1\beta_2)}}.
\end{equation}

The explicit expression for the steady states $\mathrm{T}_2$, $\mathrm{F}_1$, and $\mathrm{F}_2$ are not given here due to their excessive length. The phase boundaries in the $(\delta\alpha,S)$ parameter space can be seen in Fig. \ref{fig:phase_portraits} for both the case of higher self- and higher cross-saturation. Starting from the threshold values $\delta\alpha^{*}_{1,2}$ of the homogeneous case, we find continuously changing phase boundaries ($S^{*}_{1,2}$) to higher anisotropy values for increasing control $S$. In the case of higher self-saturation there is only one phase boundary $S^{*}_1$ due to the absence of a $\mathrm{T}_1$ state. For sufficiently strong control $S$ there is only the stable steady state $\mathrm{T}_2$. This region represents the desired outcome of a successful switching process. Its phase boundary determines the minimum control strength required to achieve switching for a given anisotropy value. For lower values of $S$, a stable four-spot pattern arises and depending on the ratio $\tfrac{\theta_1 \theta_2}{\beta_1 \beta_2}$ a second unstable four-spot pattern occurs together with a stability transition of $\mathrm{T}_2$.

\begin{figure}[t]
		\centering
		\includegraphics[width=0.48 \textwidth]{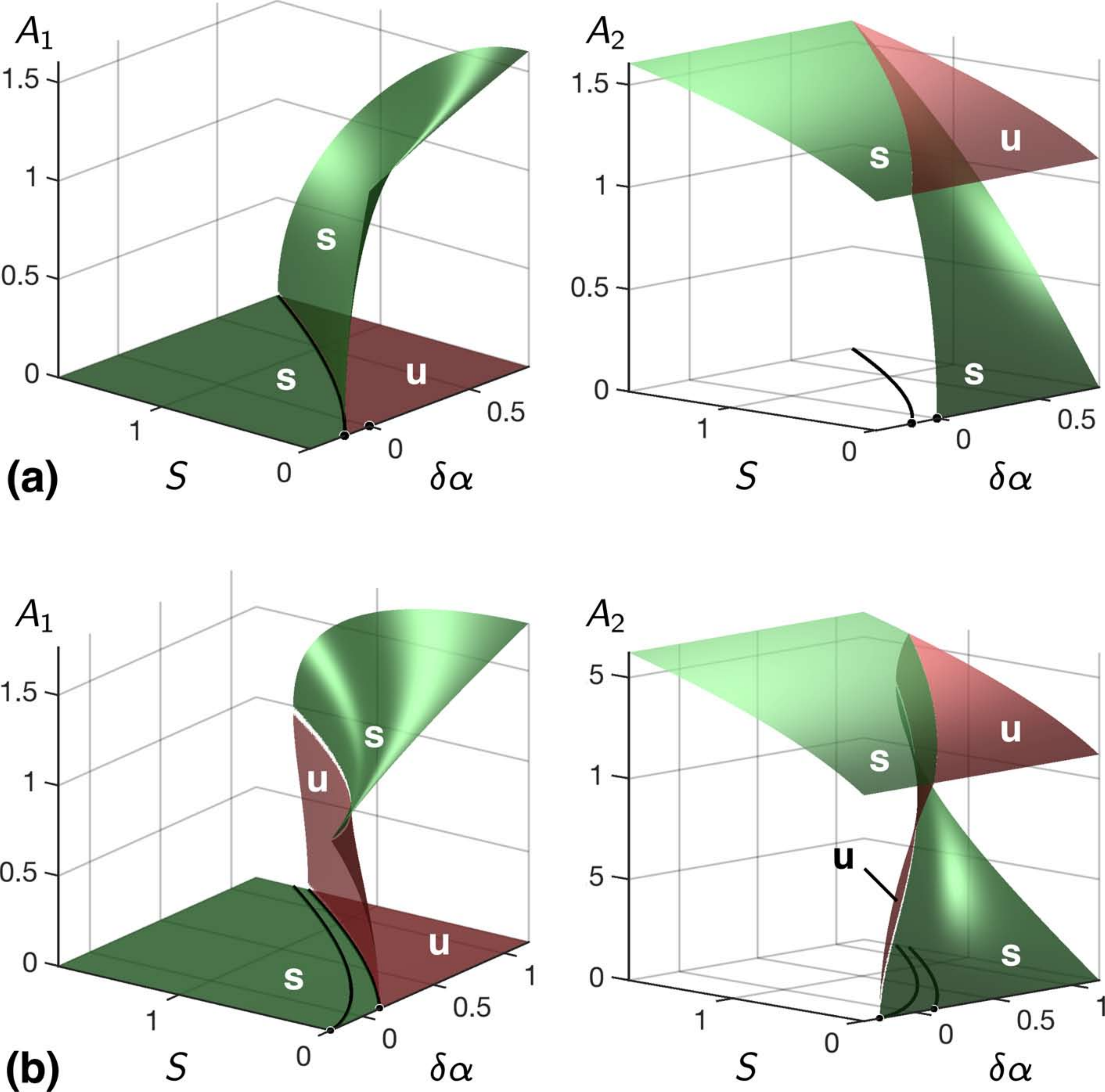}
		\caption{(Color online) Evolution of steady states for $A_1$ and $A_2$ in parameter space. \textit{Green} (s) / \textit{red} (u) surfaces belong to stable / unstable solutions. Black lines correspond to the phase boundaries projected on the $(\delta\alpha,S)$ plane (a) Larger self-saturation: one of the two pitchfork bifurcations on the $S{=}0$ line remains for $S{>}0$, whereas the other one vanishes completely. (b) Larger cross-saturation: one of the two pitchfork bifurcations on the $S{=}0$ line remains for $S{>}0$, the other one is replaced by a saddle-node (SN) bifurcation.}
		\label{fig:bif_test}
\end{figure}

Again, we can draw the flow in the $(A_1,A_2)$ phase plane and mark the basins of attraction as shown in Fig.~\ref{fig:flow}, cases 4, 5, and 6. For $\theta_1 \theta_2 {>} \beta_1 \beta_2$ a region with two attractors occurs similar to the $S{=}0$ case, but this time with a stable $\mathrm{F}_1$ state instead of $\mathrm{T}_1$, which again implies hysteresis behavior. For increasing anisotropy, the unstable $\mathrm{F}_2$ state will approach $\mathrm{T}_2$ along the unstable manifold until they meet and an unstable $\mathrm{T}_2$ and stable $\mathrm{F}_1$ state emerge. This corresponds to the approach of the two phase boundaries $S^{*}_1$ and $S^{*}_2$ and the annihilation of $S^{*}_2$, which can only happen for parameter values $1{<}\tfrac{\theta_1 \theta_2}{\beta_1 \beta_2}{<}\tfrac{3}{2}$. Otherwise the two boundaries diverge. Thus, the case of higher cross saturation is divided into two subcases, defined by either the survival or the vanishing of the middle region with two attractors.

Another important effect of the inhomogeneity $S$ is the explicit breaking of the $\mathbb{Z}_2$ symmetry in the time evolution of $A_2$. In the case of larger cross-saturation a cusp point (C) arises in the two-dimensional bifurcation diagram in Fig.~\ref{fig:phase_portraits}(b). The pitchfork bifurcation at $S{=}0$ is replaced by a saddle-node bifurcation for $S{\neq} 0$, resulting in the creation/destruction of a stable-unstable pair of F-solutions while varying one of the parameters. In the cusp point two saddle-node phase boundaries, $S_2^{*}$ and $-S_2^{*}$ (not shown), meet tangentially. The other pitchfork bifurcation ($\delta\alpha_1^{*}$) remains  ($S_1^{*}$) in presence of the inhomogeneity since the $\mathbb{Z}_2$ symmetry of $A_1$ is still conserved. This is the only bifurcation in the case of larger self-saturation in Fig.~\ref{fig:phase_portraits}(a) for $S{\neq} 0$. Figure \ref{fig:bif_test} shows the evolution of the equilibrium surfaces for $A_1$ and $A_2$ in the $(\delta\alpha,S)$ parameter space and the corresponding bifurcations. The pitchfork bifurcations are visible on the $S{=}0$ line. In the case of higher cross-saturation (lower panel) one of the pitchfork bifurcations unfolds into a saddle-node bifurcation with increasing $S$. In the case of higher self-saturation (upper panel) the supercritical pitchfork bifurcation remains stable with increasing $S$.

In conclusion, the main difference for the inhomogeneous case is the absence of a steady $\mathrm{T}_1$ state and the existence of an additional stable $\mathrm{F}_1$ state in the case of higher cross-saturation. This corresponds to the unfolding of one of the subcritical pitchfork bifurcations into a saddle-node bifurcation. The control term $S$ thus prevents extinction of $A_2$ and promotes coexistence respectively. Furthermore, the $\mathrm{F}_1$-$\mathrm{T}_2$-bistability region vanishes for high anisotropy values, if the parameter condition $1{<}\tfrac{\theta_1 \theta_2}{\beta_1 \beta_2}{<}\tfrac{3}{2}$ is satisfied, resulting in a single remaining phase boundary $S_1^*$ similar to the case of higher self-saturation. The occurring saddle-node and subcritical pitchfork bifurcations are problematic since they both imply sudden vanishing of a stable fixed point, meaning the system undergoes an abrupt transition to another stable fixed point, i.e. hysteresis is possible. This can only happen in the bistability regime (larger cross-saturation). The numerical simulation of the switching process presented in Section \ref{sec:polaritons} takes places in the larger self-saturation regime (model parameters calculated in Appendix \ref{app:sec:parameters}), which is advantageous for the switching purpose, since no hysteresis can occur. However, if the external control beam is not sufficiently strong, we observe a stable $\mathrm{F}_1$ pattern in the numerical simulations, as predicted by the PC model.

\subsection{\label{subsec:remarks}Remarks}
So far, we have discussed steady states, their stability properties, and bifurcations occurring in the solution space of the population competition model in Eq.~(2). From a dynamical perspective, we observe critical slowing down \cite{scheffer2009early,tredicce2004critical} near the phase boundaries due to the continuously vanishing real part of the Jacobian's eigenvalue responsible for the bifurcation. This corresponds to the divergence of switching times observed in the numerical simulations in Ref.~\cite{Lewandowski17}. For example, approaching $S^{*}_{1,2}$ for $\beta_1 \beta_2 {\gtrless} \theta_1 \theta_2$ from above results in divergence of the switching time. Similarly, approaching $\delta\alpha^{*}_{2,1}$ for $\beta_1 \beta_2 {\gtrless} \theta_1 \theta_2$ from the left side results in divergence of the back-switching time. Hence, to achieve favorable performance, switching should be done for parameters sufficiently far away from the phase boundaries.

We note that in general a nonlinear dynamical system can have periodic solutions which are characterized by closed orbits in the phase plane. Here, we use Dulac's Criterion \cite[p.~202]{strogatznonlinear} to rule out any periodic solutions. A simplified version reads: The existence of a function $g(A_1,A_2)$ with the property that $\nabla {\cdot} (g \mathbf{f})$ is sign definite in the entire considered state space, rules out any closed orbits in this area. If we choose $g{=}\tfrac{1}{A_1 A_2}$, we obtain $\nabla \cdot (g \mathbf{f}){=}{-}2(\beta_1 \tfrac{A_1}{A_2}+\beta_2 \tfrac{A_2}{A_1}+\tfrac{S}{A_1 A_2^2})$, and therefore closed orbits in the positive quadrant are impossible. Another observation is that for $\theta_1{=}\theta_2{\equiv}\theta$ (symmetric coupling) we can write the system in Eq.~\eqref{eq:PC model} as a gradient field $\mathbf{f}{=}\nabla V$ with the potential function
\begin{equation}\label{eq:potential}
V=\sum_{i=1}^2 \frac{\alpha_i}{2} A_i^2-\frac{\beta_i}{4} A_i^4 - \frac{\theta}{4} A_i^2 A_{i+1}^2+S_i A_i
\end{equation}
with $A_{i+2}{=}A_i$, $S_1{=}0$, and $S_2{=}S$. Closed orbits are impossible in gradient systems \cite[p.~199]{strogatznonlinear}, however, for the general case $\theta_1 {\neq} \theta_2$ (asymmetric coupling) this argument is no longer applicable. Also, assuming the gradient system defined by the potential function \eqref{eq:potential} allows us to use the language of catastrophe theory and observe that two of \textit{Thom's} seven elementary catastrophes occur, namely folds (denoted as $A_2$ in \textit{Arnold's} notation) and cusps ($A_3$). 
We further note that our detailed investigation above is limited to destabilizing linearity, stabilizing nonlinearities, and positive control parameter. These conditions match the numerical and experimental observations for the physical system under investigation here. For this case we are able to completely characterize all possible steady states, their corresponding bifurcations, and rule out any other bifurcations involving double zero and pure imaginary eigenvalues of the Jacobian.

\section{\label{sec:conclusion}Conclusions}
We have presented a detailed analysis of an all-optical switching concept based on transverse patterns in an interacting polariton system in a planar quantum-well semiconductor microcavity. From the relevant equations based on a microscopic semiconductor theory here we derived a simplified population competition (PC) model describing the system dynamics restricted to selected modes in $k$-space. Interestingly, the resulting rather simple PC model shows all key features of the system dynamics also observed in the full numerical simulations in the parameter range of interest here. In addition to what can be learned from the numerical simulations, the PC model enables us to systematically identify phase boundaries in parameter space and singularities governing the global dynamics of the nonlinear system. Interestingly the rather complicated original system of interacting microcavity polaritons, for the switching phenomenon studied here can be completely characterized by only seven flow portraits in a two-dimensional state space. The model derived has the very generic form of a generalized Lotka-Volterra (GLV) system extended with an inhomogeneity term to achieve external control. Such a system has not been investigated before. Considering the wide-spread use of GLV systems the understanding obtained in the present work is of very general nature and will be similarly applicable also to other fields where external control of population competition is studied such as chemical reactions or population competition in life-sciences.

\section{\label{sec:acknowledgements}Acknowledgments}
We gratefully acknowledge funding from the Deutsche Forschungsgemeinschaft through project SCHU 1980/5-2 and through the Heisenberg program (No. 270619725). We further acknowledge valuable discussions with Rolf Binder and a grant for computing time at the Paderborn Center for Parallel Computing ($\mathrm{PC^2}$).

\appendix
\small
\section{\label{app:sec:Numerical Details}Numerical Details}
We used the following parameters (appropriate for GaAs systems) to numerically simulate the switching presented in section \ref{sec:polaritons}: $m_{\mathrm{TE}}{=}1.05{\cdot} m_{\mathrm{TM}}{=}0.215~\mathrm{meV}\mathrm{ps}^{2} \mu\mathrm{m}^{-2}$, $\Omega{=}6.5~\mathrm{meV}$, $\gamma_c{=}0.8~\mathrm{meV}$, $\gamma_e{=}0.2~\mathrm{meV}$, $\alpha_{\mathrm{psf}}{=}5.188\cdot 10^{-4}\mu\mathrm{m}^2;$ $T^{++}{=}-5T^{+-}{=}5.69\cdot 10^{-3}\mathrm{meV}\mu\mathrm{m}^2$, $\epsilon_0^e=1.497~\mathrm{eV}$. Coherent cw excitation $2.5~\mathrm{meV}$ above the lower polariton branch was used with flat-top Gaussian profiles with intensities $I_{\mathrm{pump}}=2.1\cdot 10^{5}\cdot I_{\mathrm{probe}}=54~\mathrm{kW}\mathrm{cm}^{-2}$. The equations \eqref{eq:full model} were solved on a finite 2D grid in real space using a 4th order Runge-Kutta method with variable step size.
In Section \ref{sec:PC_model} we use parameters $\beta_1\beta_2/\theta_1\theta_2{=}1.14$ for the case of larger self-saturation and $\beta_1\beta_2/\theta_1\theta_2{=}0.69$ for the case of larger cross-saturation.
All bifurcations were determined analytically for the homogeneous case and numerically with the help of MATCONT, a matlab software package, for the inhomogeneous case.

\section{\label{app:sec:derivation}Derivation of the PC Model}
\begin{figure}[h]
		\centering
		\includegraphics[width=0.2 \textwidth]{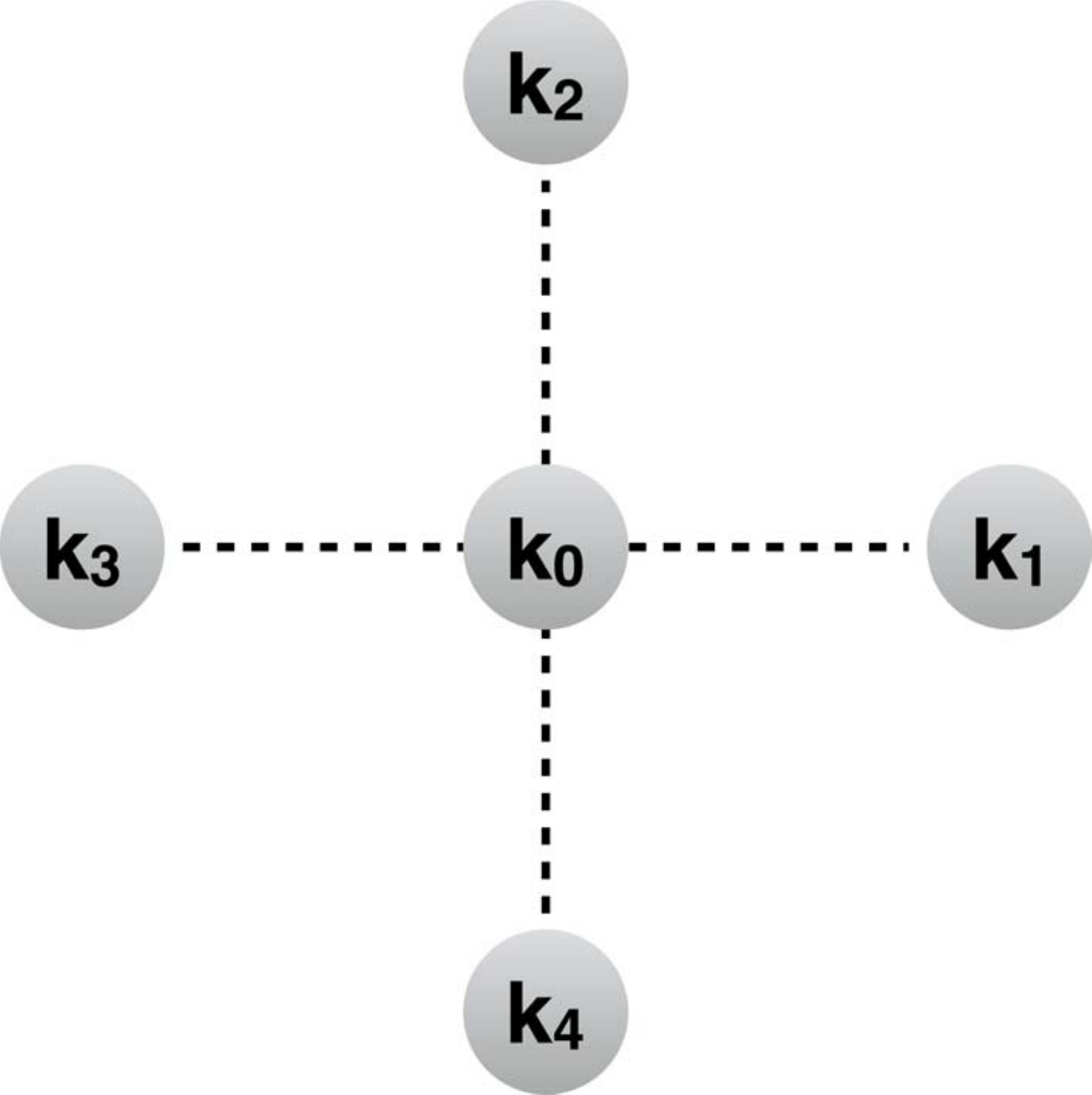}
		\caption{Definition of the reduced $k$-space.}
		\label{fig:ap:reduced k-space}
\end{figure}

We follow qualitatively the derivation of the hexagon PC model \cite{PCmodel}, but here we are including polarization effects for linearly polarized excitation, and therefore considering a different reduced $k$-space. We start with the coupled equations of motion \eqref{eq:full model} for the exciton polarization and the cavity field and transform them into $k$-space and in the linear polarization basis. Introducing the reduced $k$-space consisting of modes $\lbrace k_0{\equiv} 0, k_1,k_2,k_3,k_4 \rbrace$ with relations $k_3{=}-k_1$ and $k_4{=}-k_2$ (cf. Figure \ref{fig:ap:reduced k-space}) results in $20$ equations.
We consider an $x$-polarized pump. This leads to the following selection rules \cite{PhysRevB.94.045308}:
\begin{itemize}
\item $\substack{\mathrm{x} \\ \mathrm{y} }$-polarized probe with $\mathbf{k} \parallel \mathbf{e}_x$ excites $\substack{\mathrm{TM} \\ \mathrm{TE}}$-mode
\item $\substack{\mathrm{x}  \\ \mathrm{y} }$-polarized probe with $\mathbf{k} \perp \mathbf{e}_x$ excites $\substack{\mathrm{TE} \\ \mathrm{TM}}$-mode
\end{itemize}

We assume that all dynamical quantities oscillate with pump frequency $\omega$. Removing the phase factor $\mathrm{e}^{-\mathrm{i}\omega t}$ yields a shift of the dispersion by $-\hbar\omega$. We consider all phase-matched scattering processes uo to third order within the reduced k-space. In agreement with the linear stability analysis of the corresponding system reporting a $D_2 \cong \mathbb{Z}_2 \times \mathbb{Z}_2$ symmetry \cite{PhysRevB.77.073302}, we assume an equal excitation of opposite modes in the reduced $k$-space, i.e. $p_{\mathbf{k}_1}{=}p_{\mathbf{k}_3}{\equiv} p_1$ and $p_{\mathbf{k}_2}{=}p_{\mathbf{k}_4}{\equiv} p_2$ (analogously for $E$). We further assume the pump induced densities $E_0^{x/y}$ and $p_0^{x/y}$ at $k=0$ to be constant. Using a cyclic definition for the two mode pairs $p_j{=}p_{j+2}$ with $j{=}1,2$ results in the following phase-matched ($\mathbf{q}{=}\mathbf{k}-\mathbf{k}'-\mathbf{k}''$) scattering processes $(\mathbf{q},\mathbf{k}',\mathbf{k}'')$ for each mode pair in Eq. \eqref{eq:full model}:
\begin{equation}
(j,0,0),~ (0,j,0),~ (0,0,j),~ 3\times (j,j,j),~ 2\times (j,j+1,j+1).
\end{equation}
In our setup the pump is $x$-polarized. Therefore, terms $\propto p_0^y$ or $\propto E_0^y$ are omitted. Furthermore, we assume the co-linearly polarized off-axis modes to be very small as for these modes the instability threshold is not reached, i.e. $p_j^x \ll p_j^y$ and $E_j^x \ll E_j^y$. Hence, we also neglect terms $\propto p_j^x$ or $\propto E_j^x$. This leaves us with the following four equations for the $y$-polarized mode pairs:
\begin{widetext}
\begin{align}
\mathrm{i}\hbar\partial_t p_j^y = (\epsilon_j-\hbar\omega-\mathrm{i}\gamma_e)p_j^y+&\tfrac{1}{2}\alpha_{\mathrm{PSF}}\Omega [-p_j^{y *} p_0^x E_0^x + p_0^{x *} p_j^y E_0^x+ |p_0^x|^2 E_j^y + 3 p_j^{y *} p_j^y E_j^y + 2 p_j^{y *} p_{j+1}^y E_{j+1}^y] \nonumber \\
+ & \tfrac{1}{2}(T^{++}+T^{+-})[3 p_j^{y *} p_j^y p_j^y + 2 p_j^{y *} p_{j+1}^{y 2}] \nonumber \\
- & \tfrac{1}{2}(T^{++}-T^{+-})[p_j^{y *}p_0^{x 2}] \nonumber \\
+ & T^{++} |p_0^x|^2 p_j^y \nonumber \\
- & \Omega E_j^y \\
\mathrm{i}\hbar\partial_t E_j^y=(\hbar\omega_j^y-\hbar\omega-\mathrm{i}\gamma_c)& E_j^y -\Omega p_j^y +  E_{\mathrm{pump}, j}^y.
\end{align}

\end{widetext}
If we further assume that the time evolution of $E$ follows adiabatically the evolution of $p$ and set $\partial_t E_j {\approx} 0$, we can write 
\begin{equation}
E_j=\frac{\Omega}{\hbar\omega_j-\delta_j-\hbar\omega-\mathrm{i}\gamma_c}p_j \equiv \frac{2\lambda_j}{\alpha_{\mathrm{PSF}}\Omega} \mathrm{e}^{\mathrm{i}\theta_j}p_j
\end{equation}
with $j{=}0,1,2$. The pump field was also set to zero $E_{\mathrm{pump}, j}{=}0$ and will be added later manually. We define $\theta_j$ as the phase between $E_j^y$ and $p_j^y$ and the ratio of their amplitudes is given by $\tfrac{2\lambda_j}{\alpha_{\mathrm{PSF}}\Omega}$. Here, $\delta_j$ includes anisotropy effects due to higher density of states for the TE modes and tilting of the pump. So the parameters are given by
\begin{equation}
\lambda_j \mathrm{e}^{\mathrm{i}\theta_j}= \tfrac{\alpha_{\mathrm{PSF}}\Omega^2}{2(\hbar\omega_j^y-\delta_j-\hbar\omega-\mathrm{i}\gamma_c)}~~~~~~\mathrm{for}~j=1,2
\end{equation}
\begin{equation}
\lambda_0 \mathrm{e}^{\mathrm{i}\theta_0}= \tfrac{\alpha_{\mathrm{PSF}}\Omega^2}{2(\hbar\omega_0^x-\hbar\omega-\mathrm{i}\gamma_c)}~~~~~~\mathrm{for}~j=0.
\end{equation}
Finally, we obtain two equations for the two elementary states of the system. We also factorize the exciton field into phase and magnitude, i.e. $p_j^y{=}\tilde{p}_j^y \mathrm{e}^{\mathrm{i}\varphi_j}$, and split the results into separate equations of motion for magnitude and phase:
\begin{align}
\partial_t \tilde{p}_j^y = L_j \tilde{p}_j^y + \sum_{k=1}^2 C_{jk} \tilde{p}_k^{y 2} \tilde{p}_j^y \label{ap:eq: p} \\
\partial_t \varphi_j = K_j + \sum_{k=1}^2 D_{jk} \tilde{p}_k^{y 2}. \label{ap:eq: phi}
\end{align}
If we define $\phi_j {\equiv} \varphi_j - \varphi_0$, the coefficients are then given by
\begin{widetext}
\begin{align}
\hbar L_j  = & -\gamma_e+\lambda_0 \tilde{p}_o^{x 2} (\mathrm{sin}(\theta_0)-\mathrm{sin}(\theta_0-2\phi_j))+\lambda_j \mathrm{sin}(\theta_j)(\tilde{p}_0^{x 2}-\tfrac{2}{\alpha_{\mathrm{PSF}}}) \nonumber \\
&-\tfrac{1}{2}(T^{++}-T^{+-})\tilde{p}_0^{x 2} \mathrm{sin}(-2\phi_j) \nonumber \\
\hbar C_{jk} = & \begin{cases}
3 \lambda_j \mathrm{sin}(\theta_j)~~~~~~,~j=k \\
2 \lambda_k \mathrm{sin}(\theta_k+2\phi_k-2\phi_j)+(T^{++}+T^{+-})\mathrm{sin}(2\phi_k-2\phi_j)~~~~~~,~j \neq k
\end{cases} \nonumber \\
\hbar K_j  = & \epsilon_j-\hbar\omega - T^{++}\tilde{p}_0^{x 2}(\mathrm{cos}(\theta_0)-\mathrm{cos}(\theta_0)-2\phi_j)-\lambda_j \mathrm{cos}(\theta_j)(\tilde{p}_0^{x 2}-\tfrac{2}{\alpha_{\mathrm{PSF}}}) \nonumber \\ 
&-\tfrac{1}{2}(T^{++}-T^{+-})\tilde{p}_0^{x 2} \mathrm{cos}(-2\phi_j) \nonumber \\
\hbar D_{jk} = & \begin{cases}
-3 \lambda_j \mathrm{cos}(\theta_j)-\tfrac{3}{2}(T^{++}+T^{+-})~~~~~~,~j=k \\
-2 \lambda_k \mathrm{cos}(\theta_k+2\phi_k-2\phi_j)-(T^{++}+T^{+-})\mathrm{cos}(2\phi_k-2\phi_j)~~~~~~,~j \neq k
\end{cases}.
\end{align}
\end{widetext}
Analogously to Ref. \cite{PCmodel}, we remove the phases as dynamical variables by assuming locked phases and linearization. We define the time-dependent phase as $\phi_j (t) {\equiv} \delta \phi_j (t) + \phi_j^{(0)}$ where the locked phases satisfy $K_j(\phi_j^{(0)}){=}0$ and $\delta \phi_j (t)$ is a small deviation. Expanding equations (\ref{ap:eq: p}) and (\ref{ap:eq: phi}) up to first order in $\delta\phi_j$ around $\phi_j^{(0)}$ and neglecting terms $\propto \delta\phi_j \tilde{p}_k^{y 2}$ leads to an explicit expression for $\delta \phi_j$ which can be substituted back to obtain
\begin{widetext}
\begin{equation}\label{eq:dim_pc_model}
\partial \tilde{p}_j^y = \left[  L_j(\phi_j^{(0)}) + \sum_{k=1}^2 \left( C_{jk}(\phi_j^{(0)},\phi_k^{(0)}) - \tfrac{D_{jk}(\phi_j^{(0)},\phi_k^{(0)}) L_j'(\phi_j^{(0)})}{K_j'(\phi_j^{(0)})} \right)\tilde{p}_k^{y 2} \right] \tilde{p}_j^y.
\end{equation}
\end{widetext}
We rewrite these two equations in a shorter form:
\begin{equation}
\begin{aligned}
\partial_t \tilde{p}_1^y &=\tilde{\alpha}_1 \tilde{p}_1^y - \tilde{\beta}_1 \tilde{p}_1^{y 3} - \tilde{\theta}_1 \tilde{p}_2^{y 2} \tilde{p}_1^y \\
\partial_t \tilde{p}_2^y &=\tilde{\alpha}_2 \tilde{p}_2^y - \tilde{\beta}_2 \tilde{p}_2^{y 3} - \tilde{\theta}_2 \tilde{p}_1^{y 2} \tilde{p}_2^y
\end{aligned}
\end{equation}
We replace $\tilde{p}_j^y$ with a product of a dimensionless quantity $\hat{\tilde{p}}_j^y$ and a characteristic quantity $\tilde{p}_{j,c}^y$ which carries the original dimension, i.e. $\tilde{p}_j^y{=}\hat{\tilde{p}}_j^y \cdot \tilde{p}_{j,c}^y$. We do the same for the independent time variable $t{=}\hat{t} \cdot t_c$, so that the derivative changes to $\tfrac{\partial \tilde{p}_j^y}{\partial t}{=}\tfrac{\tilde{p}_{j,c}^y}{t_c}\tfrac{\partial \hat{\tilde{p}}_j^y}{\partial \hat{t}}$. The characteristic values are chosen in a way that the corresponding dimensionless quantities are of magnitude $1$. With the definitions $\hat{t}\equiv t$ and $\hat{\tilde{p}}_j^y \equiv A_j$, we can finally write down the population competition model as
\begin{equation}
\begin{aligned}
\partial_t A_1 &=\alpha_1 A_1 - \beta_1 A_1^3 - \theta_1 A_2^2 A_1 \\
\partial_t A_2 &=\alpha_2 A_2 - \beta_2 A_2^3 - \theta_2 A_1^2 A_2+S
\end{aligned}
\end{equation}
where we have manually added a control parameter $S$ for the $A_2$ mode pair.

\section{\label{app:sec:parameters}Calculation of Model Parameters}
The model parameters can be calculated from the physical parameters via equation \eqref{eq:dim_pc_model}. Their values, especially the signs, depend on the specific choice of the locked phases. The system's physical behavior observed in the full numerical simulations suggest destabilizing linear terms and stabilizing nonlinear terms in the PC model. Chosing phases satisfying this condition, characteristic values $\tilde{p}_{j,c}^y{=}1~\mu\mathrm{m}^{-1}$ and $t_c{=}1~\mathrm{ps}$, and anisotropy effects $\delta_1{=}0.2~\mathrm{meV}$ and $\delta_2{=}0~\mathrm{meV}$, leads to the following model parameters for the switching simulation presented in section \ref{sec:polaritons}: $\alpha_1{=}0.49$, $\alpha_2{=}0.43$, $\beta_1{=}0.007$, $\beta_2{=}0.01$, $\theta_1{=}0.006$, $\theta_2{=}0.005$. This corresponds to the case of larger self-saturation. The anisotropy is sufficiently high for the initial pattern-formation and back-switching, and also no hysteresis can occur.


\begin{thebibliography}{10}

\bibitem{Hu2008}
Xiaoyong Hu, Ping Jiang, Chengyuan Ding, Hong Yang, and Qihuang Gong.
\newblock Picosecond and low-power all-optical switching based on an organic
  photonic-bandgap microcavity.
\newblock {\em Nature Photonics}, 2:185, 2008.

\bibitem{Dawes672}
Andrew M.~C. Dawes, Lucas Illing, Susan~M. Clark, and Daniel~J. Gauthier.
\newblock All-optical switching in rubidium vapor.
\newblock {\em Science}, 308:672--674, 2005.

\bibitem{ballarini2013all}
Dario Ballarini, Milena De~Giorgi, Emiliano Cancellieri, Romuald Houdr{\'e},
  Elisabeth Giacobino, Roberto Cingolani, Alberto Bramati, Giuseppe Gigli, and
  Daniele Sanvitto.
\newblock All-optical polariton transistor.
\newblock {\em Nature communications}, 4:1778, 2013.

\bibitem{amo2010exciton}
Alberto Amo, TCH Liew, Claire Adrados, Romuald Houdr{\'e}, Elisabeth Giacobino,
  AV~Kavokin, and A~Bramati.
\newblock Exciton--polariton spin switches.
\newblock {\em Nature Photonics}, 4:361, 2010.

\bibitem{sanvitto2016road}
Daniele Sanvitto and St{\'e}phane K{\'e}na-Cohen.
\newblock The road towards polaritonic devices.
\newblock {\em Nature materials}, 15:1061, 2016.

\bibitem{Kheradmand2008}
R.~Kheradmand, M.~Sahrai, H.~Tajalli, G.~Tissoni, and L.~A. Lugiato.
\newblock All optical switching in semiconductor microresonators based on
  pattern selection.
\newblock {\em The European Physical Journal D}, 47:107--112, 2008.

\bibitem{Schumacher2009}
Schumacher Stefan, Kwong~N. H., Binder R., and Smirl~Arthur L.
\newblock Low intensity directional switching of light in semiconductor
  microcavities.
\newblock {\em physica status solidi (RRL) - Rapid Research Letters}, 3:10--12,
  2009.

\bibitem{PhysRevB.87.205307}
Luk, M. H. and Tse, Y. C. and Kwong, N. H. and Leung, P. T. and Lewandowski, P. and Binder, R. and Schumacher, S.
\newblock Transverse optical instability patterns in semiconductor microcavities: Polariton scattering and low-intensity all-optical switching.
\newblock {\em Phys. Rev. B}, 87:205307, 2013.

\bibitem{Lewandowski17}
Przemyslaw Lewandowski, Samuel M.~H. Luk, Chris K.~P. Chan, P.~T. Leung, N.~H.
  Kwong, Rolf Binder, and Stefan Schumacher.
\newblock Directional optical switching and transistor functionality using
  optical parametric oscillation in a spinor polariton fluid.
\newblock {\em Opt. Express}, 25:31056--31063, 2017.

\bibitem{cross1993}
M.~C. Cross and P.~C. Hohenberg.
\newblock Pattern formation outside of equilibrium.
\newblock {\em Rev. Mod. Phys.}, 65:851, 1993.

\bibitem{bowmannewell98}
C.~Bowman and A.C. Newell.
\newblock Natural patterns and wavelets.
\newblock {\em Reviews of Modern Physics}, 70:289 -- 302, 1998.

\bibitem{Meinhardt1982}
H.~Meinhardt.
\newblock {\em Models of Biological Pattern Formation}.
\newblock Academic Press, London, 1982.

\bibitem{murray03}
J.D. Murray.
\newblock {\em {Mathematical Biology - II: Spatial Models and Biomedical
  Applications}}.
\newblock {Springer}, {New York}, 2nd edition, 2003.

\bibitem{hassel}
M.P. Hassell, H.N. Comins, and R.~M. May.
\newblock Spatial structure and chaos in insect population dynamics.
\newblock {\em Nature}, 353:255 -- 258, 1991.

\bibitem{Kavokin}
Alexey Kavokin, Jeremy~J. Baumberg, Guillaume Malpuech, and Fabrice~P. Laussy.
\newblock {\em Microcavities}.
\newblock Oxford University Press, Inc., New York, NY, USA, 2008.

\bibitem{ardizzone}
Vincenzo Ardizzone, Przemyslaw Lewandowski, M.~H. Luk, Y.~C. Tse, N.~H. Kwong,
  Andreas L{\"u}cke, Marco Abbarchi, Emmanuel Baudin, Elisabeth Galopin,
  Jacqueline Bloch, Aristide Lemaitre, P.~T. Leung, Philippe Roussignol, Rolf
  Binder, Jerome Tignon, and Stefan Schumacher.
\newblock Formation and control of turing patterns in a coherent quantum fluid.
\newblock {\em Scientific Reports}, 3:3016, 2013.

\bibitem{egorov2014motion}
Egorov, O. A. and Werner, A. and Liew, T. C. H. and Ostrovskaya, E. A. and Lederer, F.
\newblock Motion of patterns in polariton quantum fluids with spin-orbit
  interaction.
\newblock {\em Physical Review B}, 89:235302, 2014.

\bibitem{saito2013order}
Saito, Hiroki and Aioi, Tomohiko and Kadokura, Tsuyoshi
\newblock Order-disorder oscillations in exciton-polariton superfluids.
\newblock {\em Physical review letters}, 110:026401, 2013.

\bibitem{PhysRevB.94.045308}
Lewandowski, Przemyslaw and Lafont, Ombline and Baudin, Emmanuel and Chan, Chris K. P. and Leung, P. T. and Luk, Samuel M. H. and Galopin, Elisabeth and Lema\^{\i}tre, Aristide and Bloch, Jacqueline and Tignon, Jerome and Roussignol, Philippe and Kwong, N. H. and Binder, Rolf and Schumacher, Stefan
\newblock Polarization dependence of nonlinear wave mixing of spinor polaritons
  in semiconductor microcavities.
\newblock {\em Phys. Rev. B}, 94:045308, 2016.

\bibitem{PhysRevB.77.073302}
Stefan Schumacher.
\newblock Spatial anisotropy of polariton amplification in planar semiconductor
  microcavities induced by polarization anisotropy.
\newblock {\em Phys. Rev. B}, 77:073302, 2008.

\bibitem{schumacher2007influence}
Schumacher, S. and Kwong, N. H. and Binder, R.
\newblock Influence of exciton-exciton correlations on the polarization
  characteristics of polariton amplification in semiconductor microcavities.
\newblock {\em Physical Review B}, 76(24):245324, 2007.

\bibitem{PCmodel}
Y~C Tse, Chris K~P Chan, M~H Luk, N~H Kwong, P~T Leung, R~Binder, and Stefan
  Schumacher.
\newblock A population-competition model for analyzing transverse optical
  patterns including optical control and structural anisotropy.
\newblock {\em New Journal of Physics}, 17:083054, 2015.

\bibitem{wiggins2006introduction}
S.~Wiggins.
\newblock {\em Introduction to Applied Nonlinear Dynamical Systems and Chaos}.
\newblock Texts in Applied Mathematics. Springer New York, 2006.

\bibitem{brenig1988complete}
L{\'e}on Brenig.
\newblock Complete factorisation and analytic solutions of generalized
  lotka-volterra equations.
\newblock {\em Physics Letters A}, 133:378--382, 1988.

\bibitem{hofbauer1988theory}
Josef Hofbauer, Karl Sigmund, et~al.
\newblock {\em The theory of evolution and dynamical systems: mathematical
  aspects of selection}.
\newblock 1988.

\bibitem{hernandez1997lotka}
Benito Hern{\'a}ndez-Bermejo and V{\'\i}ctor Fair{\'e}n.
\newblock Lotka-volterra representation of general nonlinear systems.
\newblock {\em Mathematical biosciences}, 140:1--32, 1997.

\bibitem{wangersky1978lotka}
Peter~J Wangersky.
\newblock Lotka-volterra population models.
\newblock {\em Annual Review of Ecology and Systematics}, 9:189--218, 1978.

\bibitem{Hering1990}
Roger~H. Hering.
\newblock Oscillations in {L}otka-{V}olterra systems of chemical reactions.
\newblock {\em Journal of Mathematical Chemistry}, 5:197--202, 1990.

\bibitem{goodwin1982growth}
Richard~M Goodwin.
\newblock A growth cycle.
\newblock In {\em Essays in economic dynamics}, pages 165--170. Springer, 1982.

\bibitem{lamb1964theory}
Willis~E Lamb~Jr.
\newblock Theory of an optical maser.
\newblock {\em Physical Review}, 134:A1429, 1964.

\bibitem{roughgarden1979theory}
Joan Roughgarden.
\newblock {\em Theory of population genetics and evolutionary ecology: an
  introduction}.
\newblock 1979.

\bibitem{Saputra2010}
Kie Van Ivanky~Saputra, Lennaert van Veen, and Gilles Reinout~Willem Quispel.
\newblock The saddle-node-transcritical bifurcation in a population model with
  constant rate harvesting.
\newblock {\em Discrete \& Continuous Dynamical Systems - B}, 14:233--250,
  2010.

\bibitem{buchberger1}
Bruno Buchberger.
\newblock {\em {E}in {A}lgorithmus zum {A}uffinden der {B}asiselemente des
  {R}estklassenringes nach einem nulldimensionalen {P}olynomideal}.
\newblock PhD thesis, University of Innsbruck, 1965.

\bibitem{buchberger2}
Bruno Buchberger.
\newblock {E}in algorithmisches {K}riterium f\"ur die {L}\"osbarkeit eines
  algebraischen {G}leichungssystems.
\newblock {\em Aequationes mathematicae}, 4:374--383, 1970.

\bibitem{scheffer2009early}
Marten Scheffer, Jordi Bascompte, William~A Brock, Victor Brovkin, Stephen~R
  Carpenter, Vasilis Dakos, Hermann Held, Egbert~H Van~Nes, Max Rietkerk, and
  George Sugihara.
\newblock Early-warning signals for critical transitions.
\newblock {\em Nature}, 461:53, 2009.

\bibitem{tredicce2004critical}
Jorge~R Tredicce, Gian~Luca Lippi, Paul Mandel, Basile Charasse, Aude
  Chevalier, and B~Picqu{\'e}.
\newblock Critical slowing down at a bifurcation.
\newblock {\em American Journal of Physics}, 72:799--809, 2004.

\bibitem{strogatznonlinear}
Steven~H. Strogatz.
\newblock {\em Nonlinear Dynamics and Chaos: With Applications to Physics,
  Biology, Chemistry and Engineering}.
\newblock Westview Press, 2000.

\end{thebibliography}

\end{document}